# On the Bistable Behavior of a Low-Frequency Raman-Active Phonon Line in Superconducting Oxide $Ba_{1-x}K_xBiO_3$

A. P. Saiko* and S. A. Markevich

*Institute of Solid-State and Semiconductor Physics, National Academy of Sciences of Belarus, Minsk, 220072 Belarus*
*e-mail: saiko@ifttp.bas-net.by*

The two-loop temperature hysteresis of integrated intensity of Raman scattering of light is explained theoretically for a line of frequency 50 cm$^{-1}$ in a superconducting oxide $Ba_{1-x}K_xBiO_3$ single crystal.

The cuprate-free superconductor $Ba_{1-x}K_xBiO_3$ (BKBO) with the superconducting transition temperature $T_c \cong 32$ K (for samples with $x \cong 0.37$) occupies a special place among high-temperature oxide superconductors. Superconducting BKBO compounds crystallize into a cubic perovskite structure, do not contain an analog of the $CuO_2$ layers inherent in cuprates, are diamagnetic, and do not possess a static magnetic order. Although the considerable isotopic effect and the superconducting gap width match the results of the standard BCS model, other mechanisms of the formation of the superconducting state (in addition to the phonon mechanism) should be employed for explaining high values of $T_c$ and low densities of states of charge carriers.

It has been generally accepted [1] that the unique properties of these compounds are determined to a considerable extent by the dynamics and the electron structure of $BiO_6$ octahedra. The vibrational spectrum of BKBO strongly depends on the dopant (potassium) concentration $x$; however, all samples with different values of $x$ exhibit a phonon vibration at a frequency of 50 cm$^{-1}$. This line appears in the Raman spectra for the $XX$ polarization owing to the $R$ type rotational distortion [2]. The line intensity is maximal for metallic samples with $x \approx 0.33$ and exhibits an intriguing feature, i.e., two-loop hysteretic behavior as a function of temperature in the range 10–200 K [2]. It is interesting to note that latticed instability with a strong softening of elastic moduli, as well as hysteretic temperature behavior for samples with $x \approx 0.37$ (which was discovered with the help of ultrasonic measurements [3]), is observed in the same temperature range.

It will be shown below that the existence of the 50-cm$^{-1}$ line in the Raman spectra and that its bistable behavior can be explained using existing concepts of the dynamics of building blocks of the BKBO crystal lattice, i.e., $BiO_6$ octahedra, as well as their electronic structure.

It was found [4] that there exist two types of $BiO_6$ octahedra with two different lengths of the Bi–O bond and rigidities. The crystal structure of the parent compound $BaBiO_3$ is formed by alteration of expanded (soft) and compressed (hard) $BiO_6$ octahedra. The electronic structures of soft and hard octahedra are also different. In a soft $BiO_6$ octahedron, 2 antibonding orbitals are completely filled with 20 electrons, while such orbitals in a hard $Bi\underline{L}^2O_6$ octahedron are filled with 18 electrons and a free energy level (hole pair $\underline{L}^2$) is present in the upper antibonding orbital. Doping of $BaBiO_3$ with potassium is equivalent to addition of holes and leads to partial replacement of large soft octahedra by small hard ones. This results in a decrease in the number of static breather and rotational distortions and their gradual disappearance. According to neutron diffraction data [5], the structure of the compound ultimately becomes, on average, cubic for a doping level of $x \geq 0.37$. However, the EXAFS analysis of the four nearest spheres in the bismuth surroundings for samples with $x \geq 0.37$ indicates local rotations of octahedra through 4°–5° [6], while Raman spectra indicate that the symmetry becomes lower as compared to simple cubic symmetry [7]. Moreover, it was found [1, 8] that oxygen ions belonging simultaneously to $BiO_6$ and $Bi\underline{L}^2O_6$ octahedra vibrate in a two-well asymmetric potential, while oxygen ions belonging to identical $Bi\underline{L}^2O_6$ octahedra vibrate in a simple harmonic potential. The two-well shape of the potential is due to different fillings of the upper antibonding orbital in $BiO_6$ and $Bi\underline{L}^2O_6$ octahedra. During tunneling of an oxygen ion from one well to the other, an electron pair simulta-neously passes from a $BiO_6$ octahedron to a $Bi\underline{L}^2O_6$ octahedron; accordingly, these octahedra exchange their roles (i.e., the dynamic exchange $BiO_6 \longleftrightarrow Bi\underline{L}^2O_6$ takes place). In [1], superconductivity in bismuth-containing compounds is regarded as movement of local electron pairs of this type.

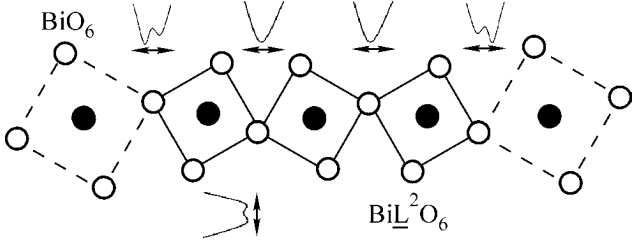

**Fig. 1.** Schematic representation of a cooperated complex of octahedra in the $BiO_2$ plane in the [100] direction (see text); dark circles correspond to Bi and light circles, to O.

With increasing doping level, the number of $Bi\underline{L}^2O_6$ octahedra increases and, for $x \geq 0.33$, the lattice acquires three-dimensional complexes, i.e., spatially overlapping $Bi\underline{L}^2O_6$ octahedra. The free energy level in these complexes expands to form the conduction band, and metal-type conductivity is formed when the energy of localization of electron pairs becomes equal to zero [1]. As a result, all octahedra in a complex can perform coherent rotational vibrations (e.g., around the [001] axis). Figure 1 shows such a complex schematically in the $BiO_2$ plane of octahedra along the [100] direction. The dynamic exchange by the roles of octahedra $BiO_6 \longleftrightarrow Bi\underline{L}^2O_6$, which is realized due to the escape of an oxygen ion from the global minimum of the asymmetric two-well potential and the transfer of an electron pair to a $Bi\underline{L}^2O_6$ octahedron, is complicated in this case since it would require mismatching of the cooperative movement of the entire complex of $Bi\underline{L}^2O_6$ octahedra. Cooperativity first leads to suppression of fluctuation-induced jumps of an oxygen ion from the global minimum to a local one upon heating of the system to a certain temperature. Second, matched rotational vibrations of the octahedra in the complex, which locally distort the cubic perovskite structure, may become Raman-active (this is actually observed for vibrations at a frequency of 50 cm$^{-1}$). The emergence of local distortions of cubic structure due to correlated rotational vibrations of octahedra in the entire complex of this type can be treated as a locally realized phase transition of displacement type (the symmetric potential with two shallow minima in the [010] direction in Fig. 1 confirms this statement).

We will consider the emergence of local distortions and lattice instability in rotational distortions of octahedra about the [001] axis by using the Hamiltonian

$$H = \sum_l \left( \frac{P_l^2}{2M} + \frac{M\nu_o^2}{2}Q_l^2 + \frac{B}{4}Q_l^4 \right) - \frac{1}{2}\sum_{l \neq l'} v_{ll'}Q_lQ_{l'}, \quad (1)$$

where $\nu_o$ is the local frequency of small oscillations in the absence of interactions ($\nu_o = 50$ cm$^{-1}$ in our case), $Q_l$ is the generalized coordinate describing the rotation of octahedra, $B$ is the anharmonicity constant, $v_{ll'}$ are the interaction constants, and $M$ is the mass. We choose the asymmetric two-well potential for describing the motion of oxygen ion belonging to $BiO_6$ and $Bi\underline{L}^2O_6$ octahedra in the form

$$V = \frac{\alpha}{2}q^2 - \frac{\beta}{3}q^3 + \frac{\gamma}{4}q^4, \quad (2)$$

where $\alpha$, $\beta$, and $\gamma$ ($>0$) are constants and $q$ is the generalized coordinate of the ion. Vibrations of octahedra depend on the position of oxygen ions in the two-well potential. This can be accounted for by introducing the cubic and quartet terms of the interaction,

$$H_{int}^{(3)} = q^2 \sum_k \lambda_k^{(3)} Q_k, \quad H_{int}^{(4)} = q^2 \sum_{k,k'} \lambda_{k,k'}^{(4)} Q_k Q_{k'}, \quad (3)$$

where $Q_k$ are the Fourier components of coordinates $Q_l$ and $\lambda_k^{(3)}$ and $\lambda_{k,k'}^{(4)}$ are the coupling coefficients. We can represent the renormalization of the initial frequency ($\nu_o \longrightarrow \nu_r$) in the self-consistent phonon approximation in the fourth-order interaction in Hamiltonian (1) and in the first and second approximations in perturbation theory in $H_{int}^{(4)}$ and $H_{int}^{(3)}$, respectively, in the form

$$\nu_r^2 = \nu_o^2 + \frac{3B}{M_{eff}}(\langle Q \rangle^2 + \langle \delta Q^2 \rangle)$$
$$+ \frac{2\lambda_{k,k'}^{(4)}}{M_{eff}}(\langle q \rangle^2 + \langle \delta q^2 \rangle) - \frac{(\lambda_k^{(3)})^2}{mM_{eff}\Omega^2}(4\langle q \rangle^2 + \langle \delta q^2 \rangle). \quad (4)$$

(see also [9],). Here, $\langle Q \rangle$ and $\langle \delta Q^2 \rangle = \langle (Q - \langle Q \rangle)^2 \rangle$ are the mean statistical values of the coordinate and its variance ($\langle q \rangle$ and $\langle \delta q^2 \rangle$ have an analogous meaning), $M_{eff}$ is the effective mass of the complex of octahedra generating local distortion $\langle Q \rangle$ (a complex may have a linear size of four to six octahedra), and $\Omega^2 = \frac{1}{m}[\alpha - 2\beta\langle q \rangle + 3\gamma(\langle q \rangle^2 + \langle \delta q^2 \rangle)]$ is the oxygen ion frequency in the two-well potential. The mean values $\langle q \rangle$ and $\langle \delta q^2 \rangle$ can be determined self-consistently from the equations

$$\langle \delta q^2 \rangle = \frac{1}{2m\Omega}\coth\frac{\Omega}{2\theta}, \quad (5)$$
$$(\beta - 3\gamma\langle q \rangle)\langle \delta q^2 \rangle = \alpha\langle q \rangle - \beta\langle q \rangle^2 + \gamma\langle q \rangle^3,$$

while $\langle Q \rangle$ and $\langle \delta Q^2 \rangle$ can be determined from the equations

$$\langle \delta Q^2 \rangle = \frac{\theta}{M_{eff}\nu_r^2}, \quad (6)$$

$$M_{eff}\nu_o^2 + 3B\langle \delta Q^2 \rangle + B\langle Q \rangle^2 - v = 0,$$

where $v = \sum_{l'} v_{ll'}$, $m$ is the mass of an oxygen ion, and $\theta = k_B T$. In deriving Eq. (4), we took into account the fact that the vibrational frequency of octahedra is much smaller than the frequency of oxygen ions. For the temperature range we are mainly interested in, we have $k_B T \gg \nu_o$; for this reason, for simplicity, variance $\langle \delta Q^2 \rangle$ is represented by the classical formula in relations (6). The value of $M_{eff}$ can be determined by comparing the local phase transition temperature $T_o$, estimated by formula $k_B T_o \approx M_{eff}\nu_o^2 \langle Q \rangle^2_{max}$, with the experimentally measured value $T_o^{exp}$. Since $\langle Q \rangle_{max} \approx 0.15$ Å for rotation of octahedra through 5° for a lattice constant $a \approx 4.3$ Å and $T_o^{exp} \approx 300$ K [2], we obtain an estimate of $M_{eff} \approx 800m$; i.e., the complex consists of 200 octahedra and occupies a region of approximately $25 \times 25 \times 25$ Å$^3$. Parameters $\alpha$, $\beta$, and $\gamma$ of potential (2) can be estimated on the basis of EXAFS data, i.e., from a potential barrier height of ~0.026 eV and the local maximum and local minimum coordinates ~0.075 and ~0.15 Å, respectively [8].

For crystals with a perovskite cubic structure, principal-order terms in the power expansion of polarizability in Raman-active modes are absent in view of the symmetry and the main contribution to the Raman scattering tensor comes from second-order terms. For this reason, we can express the integrated intensity of light scattering in the form

$$I \sim \langle Q \rangle^2 \langle \delta Q^2 \rangle = \langle Q \rangle^2 \frac{\theta}{M_{eff}\nu_r^2} \quad (7)$$

or, taking into account the second equation in (6), we can write the reduced intensity in the form

$$\frac{I}{\theta} \sim \frac{\langle Q \rangle^2}{2B\langle Q \rangle^2 + v}. \quad (8)$$

At temperatures higher than the critical temperature $T_o$ of a local phase transition of the displacement type, we have $\langle Q \rangle = 0$ and, in accordance with formula (8), intensity $I = 0$. After the emergence of local distortions $\langle Q \rangle^2 \sim T_o - T$, intensity in the temperature range $T < T_o$ differs from zero and increases upon cooling. The temperature dependence of reduced intensity $I/\theta$ calculated using formulas (8) and (4)–(6) is in qualitative agreement with the experimental curve (Fig. 2). Obviously, the temperature hysteresis in the form of a two-loop curve is due to temperature renormalization of frequency $\nu_o$ because of interactions $H_{int}^{(3)}$ and $H_{int}^{(4)}$ (see Fig. 3). It should be recalled that these interactions represent the nonlinear relations between rotational vibrations of octahedra with the breather-type vibrations of oxygen ions in a two-well potential in the [100] direction. The inclusion of the contribution from $H_{int}^{(3)}$ or from $H_{int}^{(4)}$ alone leads to a simple (one-loop) hysteresis: the intensity of scattered light in the bistability region is lower for sample cooling and higher for sample heating if $H_{int}^{(3)}$ is taken into account. If only $H_{int}^{(4)}$ is taken into consideration, the circumvention of the hysteresis curve is inverted (i.e., the cooling curve lies above the heating curve). The competition of the contributions from $H_{int}^{(3)}$ (with a minus sign, since this contribution differs from zero in the second order of perturbation theory) and from $H_{int}^{(4)}$ (with the plus sign if $\lambda_{kk}^{(4)} > 0$) forms the hysteresis curve of the two-loop type.

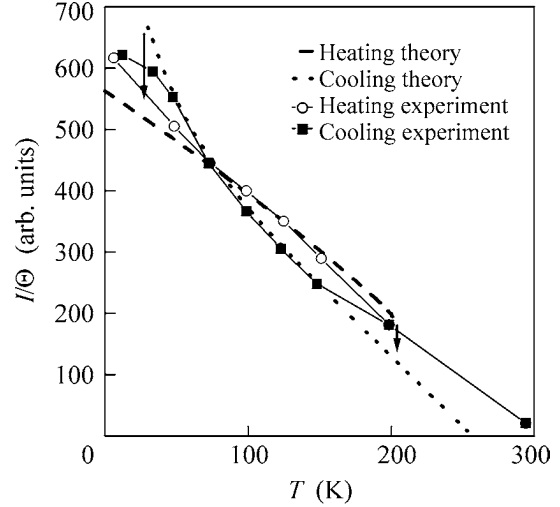

**Fig. 2.** Hysteretic behavior of the reduced integrated intensity of Raman scattering for a line at frequency 50 cm$^{-1}$ in a single crystal of Ba$_{0.67}$K$_{0.33}$BiO$_3$; $B = 4.6$ eV Å$^{-4}$, $v = 3.11$ eV Å$^{-2}$, $\lambda_k^{(3)} = 1.65$ eV Å$^{-3}$, $\lambda_{kk}^{(4)} = 1.24$ eV Å$^{-4}$, $\alpha = 3.69 \times 10^1$ eV Å$^{-2}$, $\beta = 7.5 \times 10^2$ eV Å$^{-3}$, $\gamma = 3.3 \times 10^3$ eV Å$^{-4}$, and $M_{eff} = 800m$, where $m$ is the mass of an oxygen atom.

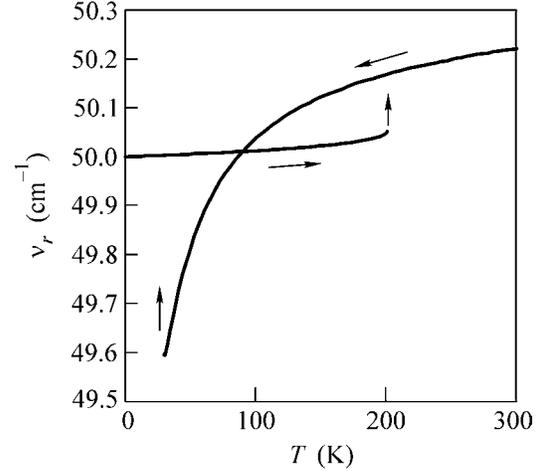

**Fig. 3.** Temperature dependence of renormalized frequency $\nu_r$ plotted in accordance with formula (4).

Thus, the presence of a phonon line of frequency 50 cm$^{-1}$ in the Raman spectrum of a single crystal of superconducting oxide $Ba_{1-x}K_xBiO_3$ and the bistable temperature dependence of its integrated intensity reflect a number of features of the local lattice and electronic structures of this compound. First, this line emerges due to rotation of soft and hard $BiO_6$ octahedra. Rotational distortions are realized in the form of a local phase transition of the displacement type in the cooperated complex of hard octahedra, which may exchange their roles with soft octahedra owing to breather-type vibrations of oxygen ions in a two-well potential. Second, the integrated intensity of Raman scattering for this line decreases upon sample heating and vanishes at the critical temperature $T_o$ of a local phase transition of the displacement type (i.e., in the case when cooperated coherent vibrations of the complex of octahedra are disrupted and local distortions in the cubic perovskite structure disappear). Third, the temperature hysteresis of the integrated intensity is due to the asymmetric nature of the dependence of rotational vibrations of octahedra with renormalized frequency $\nu_r$ on two nonequivalent positions of the oxygen ion transferring an electron pair from a soft to a hard octahedron. The bistable behavior of this ion in the two-well potential is ensured by the cooperative behavior of the complex of octahedra; in addition, the competition of the contributions from interactions $H_{int}^{(3)}$ and $H_{int}^{(4)}$ to $\nu_r$ forms a hysteretic curve of the two-loop type for the temperature dependence of the intensity.